\documentclass[onecolumn]{emulateapj} 
\usepackage{epsfig}
\usepackage{wasysym}
\usepackage[figuresleft]{rotating}
\usepackage{natbib}
\usepackage{etoolbox}
\newtoggle{emulateapj}
\toggletrue{emulateapj}

\newcommand{\msun}{\ensuremath{{\rm M}_\odot}}

\newcommand{\lsun}{\ensuremath{{\rm L}_{\odot}}}

\newcommand{\HI}{\hbox{{\rm H}\kern 0.1em{\sc i}}}
\newcommand{\HII}{\hbox{{\rm H}\kern 0.1em{\sc ii}}}
\newcommand{\CIV}{\hbox{{\rm C}\kern 0.1em{\sc iv}}}

\newcommand{\ghat}{\^G}

\newcommand{\host}{_{\rm host}}

\newcommand{\persec}{s$^{-1}$}

\newcommand{\twaste}{\hbox{\ensuremath{T_{\mbox{\rm \scriptsize waste}}}}}

\newcommand{\WISE}{{\it WISE}}

\newcommand{\kone}{\hbox{{\rm Type }\kern 0.1em{\sc i}}}
\newcommand{\ktwo}{\hbox{{\rm Type }\kern 0.1em{\sc ii}}}
\newcommand{\kthree}{\hbox{{\rm Type }\kern 0.1em{\sc iii}}}

\iftoggle{emulateapj}{\newcommand{\icarus}{Icarus}}{}
\iftoggle{emulateapj}{\newcommand{\jcap}{Journal of Cosmology and Astroparticle Physics}}{}
 \iftoggle{emulateapj}{\newcommand{\Sigurdsson}{Sigur{\scriptsize \DH}sson}}{\newcommand{\Sigurdsson}{Sigur\dh sson}}

\shorttitle{\ghat\ search for Kardashev Civilizations II}
\shortauthors{J.\ T.\ Wright et al.}
\slugcomment{Accepted for publication in {\em The Astrophysical Journal} on 23 June 2014.}

\begin{document}


\title{The \ghat\ Infrared Search for Extraterrestrial Civilizations with Large Energy Supplies. II. Framework, Strategy, and First Result}

\author{J.\ T.\ Wright\altaffilmark{1,2}, R.\ Griffith\altaffilmark{1,2}, S.\ \Sigurdsson\altaffilmark{1,2}, M.\ S.\ Povich\altaffilmark{3}, and B.\ Mullan\altaffilmark{4,5}}

\altaffiltext{1}{Department of Astronomy \& Astrophysics, 525 Davey Lab, The Pennsylvania State University, University Park, PA, 16802, USA}
\altaffiltext{2}{Center for Exoplanets and Habitable Worlds, 525 Davey Lab,  The Pennsylvania State University, University Park, PA 16802, USA}
\altaffiltext{3}{California State Polytechnic University, Pomona, Department of Physics and Astronomy, 3801 West Temple Ave, Pomona, CA 91768, USA}
\altaffiltext{4}{Blue Marble Space Institution of Science, PO Box 85561, Seattle, WA 98145-1561, USA}
\altaffiltext{5}{Carnegie Science Center, 1 Allegheny Way Pittsburgh, PA 15212, USA}

\begin{abstract}

We describe the framework and strategy of the \ghat\ infrared search for extraterrestrial civilizations with large energy supplies, which will use the wide-field infrared surveys of {\it WISE} and {\it Spitzer} to search for these civilizations' waste heat.  We develop a formalism for translating mid-infrared photometry into quantitative upper limits on extraterrestrial energy supplies.  We discuss the likely sources of false positives, how dust can and will contaminate our search, and prospects for distinguishing dust from alien waste heat.  We argue that galaxy-spanning civilizations may be easier to distinguish from natural sources than circumstellar civilizations (i.e., Dyson spheres), although {\it GAIA} will significantly improve our capability to identify the latter.  We present a “zeroth order” null result of our search based on the {\it WISE} all-sky catalog: we show, for the first time, that Kardashev \kthree\ civilizations (as Kardashev originally defined them) are very rare in the local universe.  More sophisticated searches can extend our methodology to smaller waste heat luminosities, and potentially entirely rule out (or detect) both Kardashev \kthree\ civilizations and new physics that allows for unlimited ``free'' energy generation.  

\end{abstract}


\section{Introduction and Outline}

This is the second paper in a series describing \ghat, the {\it Glimpsing Heat from Alien Technologies} survey (or G-HAT). 

In Paper I of this series, we discussed the background and justification for \ghat.  Briefly, if spacefaring extraterrestrial intelligences (ETI's) capable of colonizing other stars in its galaxy exist in the Milky Way or in another galaxy, we find, in agreement with \citet{Hart75}, that its galaxy can be fully colonized on the timescale of a galactic rotation (i.e.\ significantly less than the age of the galaxy).  Since the power generation and collection (the {\it energy supply}) of intelligent life has the potential to expand exponentially, and since intelligence implies the ability to overcome negative feedback and approach physical limits, these energy supplies could, in principle, be quite large.

An ETI's energy collection and generation must be balanced by energy disposal in steady-state.  This disposal most naturally occurs in mid-infrared wavelengths, potentially rendering large alien civilizations easily detectable and distinguishable from natural sources with new mid-infrared surveys such as that recently conducted by \WISE\ \citep{WISE}.

In order to specify and quantify the classes of ETIs that would be detectable with a waste heat search, in Section~\ref{Parameters} we develop a formalism for parameterizing the energy budget of a civilization and use it to construct models of such a civilization's spectral energy distribution (SED).  We also consider what values for the parameters we have chosen one might expect for alien civilizations, and what values are permitted by the laws of physics.

Section~\ref{WISE} describes, in broad outlines, the first aspect of our search using \WISE\ to identify stars and galaxies with anomalously high amounts of mid-infrared emission.  Here we also discuss the likely sources of false positives we will encounter and how to distinguish them from true positives.  We also present the ``zeroth order'' result of our search in this section.

We present our conclusions in Section ~\ref{conclusions}.  In subsequent papers in this series we will discuss the details of our efforts, including our quantitative results.

\section{Parameterizing Alien Civilizations}

\label{Parameters}

In this section, we develop a formalism for describing alien civilizations with large energy supplies by comparing the magnitude of their power generation and waste heat disposal to the energy available to them in starlight.  This allows us to model their SED's and associate quantitative upper limits with a null detection of waste heat.

\subsection{The Kardashev Scale}

\label{kardashev}

\citet{kardashev64} formulated a system to classify civilizations on their technological sophistication based on the scale of their energy supply (which could be used for intentional or unintentional radio communication). He defined \kone\ civilizations to be those with energy supplies comparable to humanity's supply today.  \ktwo\ civilizations had at their disposal all of the power radiated by their host star ($\sim$ 4 $\times$ 10$^{26}$ W), and a \kthree\ supercivilization could tap all of the starlight of its home galaxy ($\sim$ 4 $\times$ 10$^{37}$ W). This categorization spans an enormous range of energy supplies and conveniently divides plausible spatial extents of civilizations:  planetary, stellar system, and galactic.
  
\citet{sagan73a} noted that the gaps between Kardashev's types spanned approximately factors of $10^{10}$, and extended the definition to nonintegers by defining it in terms of the civilization's total energy supply $P$, measured in units of 10 MW:
\begin{equation}
K = \log_{10}\left(\frac{P}{10 \mbox{MW}}\right) / 10
\end{equation}

\noindent On this scale humanity's current energy supply makes it a $K=0.7$ civilization, and all life on Earth has $K=0.89$ \citep{hierarchy}.

In this work, we follow \citet{Zubrin00} and apply Kardashev's types more loosely, defining a \kone\ civilization (which we will distinguish from Kardashev's definition with the notation {\it K1}) as one that ``has achieved full mastery of all its planet's resources''.  That is, we do not tie it specifically to an absolute energy supply (since other planets may have different capacities of power) but rather to the extent and approximate technological achievement of the civilization.  By \citeauthor{Zubrin00}'s definition, humanity is nearly a K1 civilization.  Analogously, a $K2$ civilization is a circumstellar civilization, tapping a significant fraction of its host star's light and having the capacity for resource acquisition and use that far exceeds that of a planet-bound civilization.  Finally, $K3$ refers to a supercivilization that is galaxy-spanning, composed of so many K2's that it has mastered a significant fraction of the galaxy's luminosity and resources.

\subsection{The AGENT Formalism}

\label{AGENT}

We describe here a general formalism for parameterizing the effects of alien waste heat on the spectrum of a star or galaxy.  This formalism is general in the sense that it attempts to parameterize an alien civilization's energy budget without making any assumptions about the nature of alien civilizations or their engineering.  

We divide energy supply (that is, generation and collection) into two components:  starlight collection and other sources, which we specify with the symbols $\alpha$ (which is meant, for mnemonic purposes, to remind the reader of an absorption coefficient) and $\epsilon$ (meant to remind the reader of the energy generation per unit volume in stellar structure equations).  We likewise divide energy emission or disposal into two categories: waste heat in the form of radiation (denoted $\gamma$, meant to evoke a common symbol for the photon) and all other forms (denoted $\nu$, meant to suggest neutrino radiation, an example of a source of non-thermal or non-photon energy losses, as in, for example, supernovae).   Finally, where appropriate, we assume that thermal waste heat has some characteristic temperature \twaste, which may be a function of the portion of the electromagnetic spectrum observed.  

While these parameters could be expressed in physical units, we find it more convenient to normalize them by the energy available to a civilization in starlight.  Thus, we normalize by the incident starlight on a planet for a K1, by a stellar luminosity for a K2, and by a galactic luminosity for a K3.  This restricts $\alpha$ to lie between zero and one, but the other parameters can take any positive value.
\begin{table}
\begin{center}
\begin{tabular}{c p{2in} c p{2in}}
\multicolumn{4}{c}{{\sc Table~1: Definition of parameters in the AGENT formalism, and values for humanity}}\\
\hline 
\hline
{Parameter} & {Meaning (Powers as fractions of available starlight)} & {Value for Humanity} & {Notes} \\
\hline
$\alpha$ & Power of intercepted starlight &        $\sim 10^{-7}$ & Counting only photovoltaic generation (ignoring passive heating, agriculture, etc.) \\  
$\epsilon$ & Power of non-starlight energy supply (e.g.\ fossil fuel use, zero-point energy, nuclear energy) &      $\sim 10^{-4}$& Mostly fossil fuel, some nuclear \\ 
$\gamma$ & Power of waste heat in the form of thermal radiation of photons      &    $\sim 10^{-4}$ & $\sim  \epsilon$ \\     
$\nu$ & Power of other waste disposal (e.g.\ neutrino radiation, non-thermal emission, kinetic energy, energy-to-mass conversion)&     negligible & Radio transmission and radar, neutrino losses in nuclear reactors, kinetic and potential energy in spacecraft\\                       
\twaste & Characteristic color temperature of thermal waste photons     &   $\sim 285$ K & Typical operating / ambient temperature \\
\hline
\end{tabular}
\end{center}
\end{table}

We will refer to this as the {\it AGENT} parameterization, after the parameters $\alpha\gamma\epsilon\nu\twaste$, and present the parameters in Table~1.

Conservation of energy and steady-state together require that the energy a civilization collects or generates must equal the energy it emits or disposes of, most naturally as waste heat.   This yields:

\begin{equation}
\alpha+\epsilon = \gamma+\nu
\end{equation}

The assumption of negligible non-thermal losses ($\nu\sim 0$) and the further assumption that all sources of power other than starlight are negligible ($\epsilon \sim 0$) yields the useful simplification of $\alpha = \gamma$.

To illustrate the AGENT parameters, consider that humanity has an energy supply of approximately $1.4 \times 10^5$ TWh per year, or approximately $1.7 \times 10^7$ MW, mostly from fossil fuels \citep{hierarchy}.  Since the total amount of sunlight incident on the Earth is $1.7\times 10^{11}$ MW, we parameterize humanity as having values given in Table~1.

We have made several simplifying assumptions for this calculation.  We have ignored sunlight used for passive heating, agriculture, and the many other ways it makes Earth inhabitable for us, because we are here isolating our civilization's contribution to the global energy budget.  We have also assumed steady state, and so ignored forms of energy storage such as the potential energy in architecture, the chemical energy in stored food and batteries, and similar considerations.  Fossil fuels are stored solar energy, but the time over which this storage took place is so long that their use does not badly violate our steady-state assumption.

Put this way, we see that humanity warms the Earth with its energy generation by 0.01\% above the effects of sunlight alone, and this effect has elevated the Earth's MIR heat signature by that amount (0.1 millimagnitude) above its pre-industrial levels.   This effect (direct heating) is currently small compared to greenhouse warming (which insulates the Earth's surface but does not change the global energy balance) and albedo effects (such as loss of reflective sea ice or changes in cloud cover).  

The energy budget of life on Earth is surprisingly poorly known, but on average photosynthetic life on Earth generates $\sim 2\times10^8$ MW from sunlight with approximately $2\%$ efficiency \citep{hierarchy}, implying that it ultimately intercepts $\sim 5\%$ of sunlight that strikes the top of the atmosphere.  Since photosynthetic energy production dominates Earth's biological work budget, we would write that for all life on Earth $\alpha = 0.05$ and, since all of this energy eventually cascades into waste heat, $\gamma = 0.05$.

\subsection{Illustrative Examples}

We now illustrate our formalism by translating some speculative forms of advanced civilizations from the scientific and popular literature into the AGENT parameterization.

First, we consider civilizations with energy supplies comparable to their star's total luminosity.  A ``classical'' Dyson sphere that generated power solely from its completely enfolded solar-luminosity star and with a radius of 1 AU would have $\alpha = \gamma = 1$; $\twaste \sim 250 $ K; and $\nu = \epsilon = 0$.  A civilization using a rotating black hole to convert one Jupiter mass of material to energy with 40\% efficiency evenly over the course of 5 Gyr via the Penrose mechanism would have a energy supply of approximately 1 solar luminosity, so $\epsilon = 1$, but might have  $\alpha = 0$ if it employed no mega-engineering to block significant amounts of starlight.  A ``ring world'' \citep{Ringworld} with radius $r=1$ AU and with a width of $w=1.6 \times 10^6$ km would intercept $\frac{1}{2} w/r$ of its star's light, and so have $\alpha = 5\times10^{-4}$.  

For K3's, $\alpha$ would represent the luminosity-averaged covering fraction of stars in the galaxy, so $\alpha = 1$ corresponds to the limiting case of every star being enshrouded by a complete Dyson sphere, and $\alpha = 0.5$ could represent partial coverage of every star or total coverage of stars contributing one half of the galaxy's total luminosity.  

We can also translate some past SETI work into the AGENT parameters.

Radio or optical communication SETI can be thought of as searches for the deliberate or leaked power we parameterize by $\nu$.  In Kardashev's original formulation of types of civilizations, he made an order of magnitude estimate of the power available for (non-thermal) radio emission from civilizations  given that it must be some fraction of the energy supply.  In our parameterization Kardashev estimated $\nu \lesssim (\alpha+\epsilon) \approx 1$.   A variation on this theme is the idea of \citet{Loeb11} to search for leaked high brightness-temperature light from alien civilizations on Kuiper Belt Objects.

One can approximate a civilization's place on Sagan's version of the Kardashev scale (\S~\ref{kardashev}) for galaxies similar to the Milky Way and stars similar to the Sun as 

\begin{equation}
K = [K] + \log_{10}(\alpha+\epsilon)/10
\end{equation}

\noindent where $[K]$ is the ordinal Kardashev type on Zubrin's scale.  Thus, a civilization that is intercepting 5\% of its Sun-like host star's light and producing another 0.05 \lsun\ from other sources around its planetary system would be a K2 (it is circumstellar) with approximately $K=2 + \log_{10}(0.05+0.05)/10 = 1.9$ (the proper value in Sagan's formulation is $K = \log_{10}(0.1 \times 4\times10^{20})/10 = 1.96$).

The K3 search of \citet{annis99b} sought evidence of stellar obscuration on the basis of optically faint galaxies that appeared to be outliers to the Tully--Fischer relation, and was sensitive to galaxies with greater than 75\% obscuration.  In AGENT parameters, Annis sought galaxies with $0.75 < \alpha < 1$ (although his sensitivity near 1 was necessarily limited by the fact that the galaxies were required to have a measured optical rotation curve to be included in his sample, creating a selection bias against the highest-$\alpha$ galaxies).  

We can contrast this with the MIR search originally proposed by Dyson, which is focused on  high-$\gamma$ objects.   This describes the searches already performed by researchers such as Jugaku and Carrigan, and such as we will describe later.  Indeed, much of the work of Jugaku put rather tight limits on the values of $\gamma$ for nearby stars with {\it IRAS}, as low as $\gamma < 0.01$.

The artifacts postulated by \citet{Arnold05} (see \S~2.2 of Paper I) would correspond in AGENT parameters to structures with $\gamma = \alpha \sim 10^{-2}$--$10^{-4}$, with the twist that the detection and communication would be made through the time variations in $\alpha$.

Viewed this way, the principal advantage to searches for electromagnetically telecommunicating civilizations \citep[i.e, searches for high-$\nu$ civilizations,][]{blair92} is that the power emitted might be in a narrow bandwidth (and so have high flux density), might be obviously of alien origin, and might thus be easily distinguished from background or natural sources of radiation.  The principal disadvantage is that we must guess at the form of $\nu$ and/or hope it is large enough for us to detect with our technology and from our vantage.  In contrast, the principal advantage of searches for waste heat is that we expect $\gamma \gg \nu$, and we expect the waste heat to have a characteristic (thermal) spectrum; their principal disadvantages are that they are only sensitive to large values of $\gamma$, and that there are many confounding sources (mostly involving stochastically heated dust) that will mimic this signature.   Thus, the two forms of SETI are most sensitive to different kinds of activities of alien civilizations and face different challenges.

\subsection{Spectral Implications}
\label{spectra}

We can use the AGENT parameters to roughly calculate the MIR spectrum we would expect for large civilizations.  

Consider a simple model in which a galaxy or star at distance $d$ has a stellar luminosity $L\host$ and a flux density received at Earth $F_\lambda$.  We describe the SED of the host galaxy, $f\host$, normalized to unity area, i.e.
\begin{equation}
\lambda F_\lambda  = f\host \frac{L\host}{4\pi d^2}
\end{equation}

This galaxy hosts a civilization with typical waste heat temperature \twaste\ blocking a fraction $\alpha$ of the stellar luminosity and generating non-thermal emission with normalized SED $f_{\rm nt}$.

The presence of the civilization will, to first order, alter the spectrum received at Earth to be:

\begin{equation}
\label{alphaeq}  
\lambda F_\lambda = \left[(1-\alpha) f\host + \gamma \frac{\lambda \pi B_\lambda(\twaste)}{\sigma \twaste^4} + \nu f_{\rm nt} \right] \frac{L\host}{4\pi d^2}
\end{equation}

\noindent where $B_\lambda(\twaste)$ is the specific intensity of the Planck function at $T=\twaste$.  Note that this expression ignores any feedback on the host stars from the presence of a large solid angle of collectors \citep[e.g.][]{badescu2000}, and any non-stellar sources of radiation, such as the emission from dust or a galaxy's central engine.  We have also assumed in this equation that all non-thermal energy disposal (represented by $\nu$) is electromagnetic (i.e.\ it can be represented by $f_{\rm nt}$).

For K2's, the starlight may not be blocked uniformly at all angles (that is, modeling the stellar energy collectors as a translucent screen may be inappropriate).  If we assume that the stellar disk is completely unocculted at most times, then we can ignore the $(1-\alpha)$ factor.  For K3's, the number of stars in the galaxy makes it appropriate to take a phase average over them and retain the $(1-\alpha)$ term.  

\subsection{Simple Blackbody Model}

\label{Toy}

If we crudely approximate a stellar or galactic spectrum as a blackbody with temperature $T_*$ and ignore $\nu$, then we can rewrite the expected SED as

\begin{equation}
F_\lambda \approx \left[(1-\alpha) B_\lambda(T_*) + \gamma B_\lambda(\twaste) \left(\frac{T_*}{\twaste}\right)^4\right]\frac{\pi}{\sigma T_*^4}\frac{L\host}{4\pi d^2}
\end{equation}

\noindent If, more realistically, we restrict this last approximation to the Rayleigh--Jeans limit, we have
\begin{eqnarray} 
F_\lambda (\lambda \gg hc/(kT_*)) & \approx &  \left[1-\alpha + \gamma \left(\frac{T_*}{\twaste}\right)^3\right]\frac{ck}{2\lambda^4\sigma T_*^3}\frac{L\host} {d^2} \\
& \approx & \left[1-\alpha + \gamma \left(\frac{T_*}{\twaste}\right)^3\right]F_\lambda (\alpha=\gamma=0) 
\end{eqnarray}

\noindent  This form reveals the well-established power of working in the MIR when searching for high-$\gamma$ civilizations.  We see that the MIR flux of high-$\gamma$ civilizations will be strongly enhanced by the large $(T_*/\twaste)^3$ term, which will be on the order of $10^3$ {\it even for modest values of $\gamma$}.  Indeed, for $\gamma=0.15$, $T_*=4000$ K, and $\twaste=255$ K, this amounts to an increase in MIR flux of nearly {\it 7 mag}.   Further, the MIR color temperature would be dominated by the 255 K waste heat, meaning that it will be distinguished from most astrophysical sources in both luminosity and color.  By contrast, the dimming in the NIR/optical is linear in $\alpha$, and so is a much smaller effect, only observationally distinguishable for values of $\alpha$ near 1.

The lesson here is that to have detectable waste heat, a civilization need not completely obfuscate its star(s) nor must it employ starlight as the primary source of its energy.  The MIR excess scales, to first order, as $\gamma (T_*/\twaste)^3$, which is large enough that the primary obstacle to detecting waste heat from high-$\gamma$ civilizations will not be their brightness, but confusion with other astrophysical sources with similar broadband colors or spectra.

\subsection{Plausible Values for the AGENT Parameters}

\subsubsection{Starlight Absorbed: $\alpha$}

\label{Esources}

In the solar system, the available free energy is dominated by the Sun's luminosity, which provides $\sim 10^{33}$ erg \persec\ for $\sim 10^{10}$yr, or $\sim 10^{51}$ erg over its lifetime, having converted $10^{-3}$ of its rest mass into energy.    The planets collectively contain $\sim 10^{-3}$ \msun, meaning that even perfectly efficient mass-to-energy conversion would provide no more energy than the Sun provides ``for free'' if performed over the lifetime of a star.   The gravitational potential and kinetic energy of the planets are both dominated by Jupiter, and are $\sim G\msun M_{\mbox \jupiter} / (5 {\rm AU}) \sim 10^{42}$ erg.  Coincidentally, this is similar to the rough order of magnitude of the chemical potential energy of the planets ($\sim M_{\mbox \jupiter} / m_{\rm p}$ eV $\sim 10^{43}$ erg; i.e.\ the energy available for ``burning'' them).

We see, then, that the Sun dominates the energy budget of the solar system both in terms of readily available energy and in mass available for both construction and mass-to-energy conversion.  Of course, a civilization might use the mass of planets as an energy source for timescales much less than 10 Gyr, in which case its energy supply might temporarily exceed that available from its host star.  Such a energy supply would necessarily be short-lived, however, and so the chances of its detection would be quite small (given the age of the Galaxy and the duty cycle this implies).  Long-lived civilizations with large energy supplies might therefore be expected to rely almost entirely upon starlight for their energy needs.

Thus, for civilizations relying primarily on starlight for their power, collected starlight equals power disposal ($\alpha = \gamma + \nu$), so power disposal will be some fraction of total stellar luminosity ($0 < (\gamma+\nu) < 1$), and other power sources are negligible ($\epsilon \ll 1$).

\subsubsection{Other Energy Sources: $\epsilon$}

\label{epsilon}

An arbitrarily advanced civilization may, of course, be able to tap additional sources of energy.  Speculation into the nature of such technology is perilous, but we can explore the limitations imposed on it by our current understanding of the laws of physics.   The conversion of mass to energy through fusion or fission is generally inefficient (and, at any rate, stars already do this ``for free''), but in theory the use of a rotating black hole (through the Penrose mechanism) can provide very high efficiencies \citep{PenroseEfficient}.    The star itself might provide the fuel for such an engine.  If such a process operated at 40\% efficiency and consumed the entire star over its natural lifetime, it would effectively increase the luminosity of a star system by a factor of $(0.4/0.001) = 4\times10^4$ (if the higher luminosity were to be sustained over 10 Gyr).

K3's would also have two additional resources at their disposal:  interstellar gas and the central supermassive black hole.  The former should be of little utility:  the gas in a galaxy typically composes only 10\% of its baryonic mass, and is inconveniently distributed compared with the mass in stars (in the sense that it is not dense, and so must be collected across large volumes to be of much use).  The central black hole might provide the core of an engine for converting, for instance, stellar mass into free energy \citep{Inoue11};  indeed such an engine might bear a striking similarity to active galactic nuclei (AGNs).\footnote{The idea that {\em all or most} AGN are, in fact, alien engines that we have misinterpreted as natural phenomena has a certain ironic appeal, and might explain away any theoretical difficulties in perfectly modeling them (i.e.\ what appears to be missing physics is in reality deliberate engineering on the part of alien civilizations).  Besides being an ``aliens of the gaps'' argument, though, this explanation fails to explain the abundance of AGN with cosmic time (i.e.\ why AGN are most abundant at redshift and relatively rare in the local universe), since one would na\"ively expect their abundance to increase with time as K3's arise and reach their central engines.  That is, it is unclear why alien civilizations would almost invariably shut off their engines after a few gigayears.  This explanation would also require that K3's arise very quickly after the Big Bang (even at $z>5$) which seems quite unlikely.}
 
Since dark matter dominates the mass-energy budget of a galaxy (composing, e.g., 90\% of the Milky Way's mass), it is in some sense the most obvious and abundant energy source, and bears consideration as a source of energy.  The most widely accepted model of cosmology holds that $\sim 85\%$ of the mass-energy of the universe (not counting the ``dark energy'' responsible for cosmic acceleration) is in the form of dark matter.  This matter is cold (in the sense that the kinetic energy per particle $kT\ll m c^2$) and, so far, is detected only by its gravity.  The most popular candidate for the dark matter is an as-yet undiscovered, relic, weakly interacting, massive particle.  The dark matter is known to be self-gravitating and bound down to at least the scales of galaxies, and galaxy clusters host high concentrations of dark matter, especially at their centers.  The details of the temperature and nature of the dark matter are still poorly constrained observationally, but this outline is sufficient to consider its applicability to the problem of alien energy supplies.

If we very roughly take a large galactic dark matter halo to be $M\sim 10^{12}$ solar masses ($10^{45}$ g) with typical size $R\sim 10^{23}$ cm ($\sim 30$ kpc), then the average mass-energy density is $\sim 10^{-24}$ g/cm$^3 \sim 10^{-4}$J/m$^{-3}\sim 0.6 $GeV/c$^2$/cm$^{3}$.  The typical velocity of the dark matter particle or clump will be $\sim \sqrt(GM/R) \sim 300$ km/s, so the average mass-energy flux would be of the order of $10^2$ W/m$^2$.  This is roughly one-tenth of the solar constant of 1400 W/m$^2$ (comparable to the output of a modern photovoltaic solar cell) and so would warm a surface to $\sim 200$K. 

So in principle, especially in dark-matter halos with deep gravitational wells and large particle fluxes, dark matter may be a useful source of energy over gigayear timescale, if it could be tapped with high efficiency.  The major obstacle to dark matter energy extraction is that most dark matter candidate particles have nearly negligible cross sections to baryonic matter and electrons, and so cannot be captured or otherwise used as a source of energy with high efficiency.  If the dark matter particles have cross sections to baryonic interaction similar to that of the neutrino, then it would require neutron-star-like densities to achieve optical depths approaching unity on scales of 1 m (but, since neutron matter is degenerate at reasonable temperatures, the particles still could not interact with the target through elastic scattering because all of the possible post-interaction states for the target particle would be filled).  Alternatively, gravitational focusing might be employed to create especially high densities of dark matter, which might self-annihilate if it is composed of both particles and their antiparticles, and might decay into a form useful for energy extraction.\footnote{An interesting consequence of such an energy source is that since the gravity of dark matter holds galaxies together, its conversion into thermal photons would slowly unbind galaxies and galaxy clusters.  The environmental consequences of depleting this energy source would be quite dramatic.}   Indeed, \citet{Hooper12} suggested dark matter as a potential source of internal heating for planets that might make them habitable (i.e.\ capable of supporting liquid surface water).

Regardless of the form overcoming this obstacle might take, it would imply that a K3 could generate a modest amount of energy anywhere within a dark matter halo (although it is not clear what benefit this would provide over a small nuclear reactor other than the potential for saving the necessary reactants over billions of years).  

Even more exotically, alien civilizations might be able to tap zero-point energy or cosmic dark energy for their needs.  It is unclear to us to what degree this is allowed or prohibited by the laws of physics.
 
Regardless of the form of non-starlight energy an ETI might take, it must be ubiquitous across a supercivilization for it to contribute significantly to a galactic energy budget, and so this alternative energy source must be ``obvious'' or easily replicable.  If such energy sources are available, then they might supersede starlight as the dominant energy source in the galaxy.  In this case, we might expect $\epsilon \gg \alpha$, and even $\epsilon \gg 1$.  If this energy were to be emitted as thermal waste heat  (i.e.\ if $\epsilon \sim \gamma$) at high waste heat temperatures, then such a civilization might be quite easily detected in the MIR, but hardly noticeable at all by way of obscured starlight.  Indeed, such a K3 would be so noticeable that even a cursory examination of the \WISE\ survey catalog shows that they must not be common in the local universe.

\subsubsection{Power Disposal: $\gamma$ and $\nu$}

\label{nu}
The emergent spectrum of an alien civilization will depend on the sort of work it does with its energy supply.  On Earth, most of the work we do eventually results in little to no energy storage: transportation, metabolism, energy distribution, and computation all accomplish useful tasks, but essentially all of the energy involved is ultimately dissipated as waste heat.  We might then expect that alien civilizations will do the same, but we must also account for other possibilities.

The most obvious sources of non-thermal disposal of energy are radio emission and other electromagnetic radiation.  On Earth, small amounts of our energy supply is lost to space in these forms as radar, high brightness temperature outdoor lighting, laser, and other decidedly non-thermal radiation.  Searches for alien analogs to this emission compose significant parts of optical and radio communication SETI.  Other possibilities for non-thermal energy use are energy-to-matter conversion, or the generation of non-E-M forms of radiation (gravity waves, neutrinos, etc.).  These all represent low-entropy forms of energy, and the second law of thermodynamics requires that the balance of entropy be carried away by waste heat.

\citet{slysh85} noted that the maximum possible thermodynamic efficiency of a civilization is the Carnot efficiency 
\begin{equation}
\eta = 1-\twaste/T
\end{equation} 
\noindent where $T = T_*$ for collected starlight and is characteristic temperature at which energy is generated otherwise.  We can use this observation to calculate an upper limit on $\nu$.  If a civilization operates at maximum thermodynamic efficiency, then the fraction of energy $\nu$ that does not ultimately end up radiated away as waste heat (i.e. remains at low entropy) is given by 
\begin{equation} 
\frac{\nu}{\alpha+\epsilon} \leq \eta
\end{equation} 
For a fiducial Dyson sphere at 1AU around a Sun-like star, we have $\eta \approx 1-290{\rm K} /5800{\rm K} = 0.95$, so in this case $\gamma / \alpha \ge 0.05$, $\nu/\alpha \le 0.95$, and $\nu/\gamma \le 19$.  

More exotically, it is conceivable that ETI's might employ machinery that is based on the weak force, and thus radiate significant waste heat as thermal neutrinos, or other particles that do not quickly decay into thermal photons, in which case they might be able to have $\nu \gg \gamma$.  If such physics is obvious and pervasive enough to be employed by all ETI's, we would not expect to be able to detect them from their waste heat in photons.

If a civilization's waste heat from stellar energy is typically photonic then $\gamma/\alpha \ge \twaste/T_*$ and, by conservation of energy, $\nu = \alpha + \epsilon - \gamma$.  If, like humanity, ETI's do most of their work in a manner that eventually generates waste heat ($\gamma \sim \alpha + \epsilon$) then waste heat will dominate the civilization's radiation ($\nu \ll \gamma$).  If, further, starlight dominates a civilization's energy supply, then we have ($0\lesssim\gamma\lesssim1$).

\subsubsection{Waste Heat Temperature: $\twaste$}

\label{twaste}

We expect \twaste\ to have values near 300 K, but not solely for anthropocentric reasons.  Of course, since Earth demonstrates that intelligent, spacefaring life can emerge and thrive at these temperatures, it follows that we should expect at least some alien civilizations to favor this temperature.  However, the waste heat temperature does not need correspond to either preferred ambient temperature of alien organisms or to the effective temperature of the collectors of stellar power; it represents the temperature of the outermost radiating surface of a stellar system.  Nonetheless, there are thermodynamic reasons that we do not expect waste heat to be disposed of at temperatures far from 300 K for civilizations using starlight.

The waste heat from the outermost radiating surfaces of a civilization could itself be used to do work at a yet lower temperature.  Thermodynamically, this corresponds to an improvement in the maximum possible efficiency, $\eta$ from starlight at a given temperature.  Naively, it would seem then that ETIs with large energy supplies would construct highly efficient Dyson spheres, potentially rendering waste heat so cool that it would not radiate in the MIR.

However, the marginal free energy yield from employing ever lower waste heat temperature decreases with \twaste\, even as the difficulty of the engineering rapidly increases.  Energy balance for a Dyson sphere with $\epsilon=\nu=0$ requires that
\begin{eqnarray}
  L_* \alpha = \gamma \sigma\twaste^4 A
\end{eqnarray}
\noindent where $A$ is the surface area of the radiators.  Thus, for $\alpha = \gamma$, the radiating surface $A$ must increase as $(1/\twaste)^{4}$.   For instance, improving the maximum thermodynamic efficiency $\eta$ of a K2 from 95\% (for instance, a 1 AU sphere radiating 290 K waste heat reprocessed from 5800 K starlight) to 99.5\% (the same, but radiating at 29 K) would require an increase in surface area of 
\begin{equation}
\left(\frac{1-0.95}{1-0.995}\right)^{-4} = \left(\frac{29{\rm K}}{290 {\rm K}}\right)^{-4} = 10^4
\end{equation}
\noindent(representing a sphere 100 AU in radius in our example).  

This strong inverse dependence implies that there will be an optimal waste heat temperature beyond which it would be impractical to extract more free energy from starlight.   This optimal operating temperature would then characterize Dyson spheres, and this temperature will likely be significantly above the minimum possible (i.e.\ that of the cosmic microwave background at 2.7 K).    Indeed, the $\twaste^{-4}$ dependence is so strong, one might expect that civilizations will operate at the highest temperature consistent with the work they wish to do.

Of course, different civilizations might optimize this problem in different ways, but there are reasons to expect a value near 300 K.  At this temperature around Sun-like stars, $\eta = 95\%$, a point at which there is an order of magnitude less free energy available in the waste heat as there is in the starlight itself.  Further improvements in maximum thermodynamic efficiency at this point are small compared to the potential value of improving the energy efficiency of the ``factories'' themselves.  This temperature is also low enough that complex molecules can form and most rocks and metals are solid, allowing for the construction of mega-engineering projects.  Temperatures between 150 and 600 K are thus ``reasonable'' values for the optimization of this problem (spanning a factor of 250 in radiating surface to be engineered).

As long as this optimal $\twaste$ value is measured in this range, then the Wien peak of a civilization's waste heat --- and thus the bulk of its luminosity --- should be in the MIR, and we should expect searches there to be effective.  This is in contrast to natural astrophysical sources of radiation with significant amounts of dust, which can have much lower effective temperatures (because dust clouds are large, having very large effective radiating area, their typical \twaste\ values are 15--50 K).
 
The highest thermodynamic efficiencies for K2's will result in waste heat having an SED very close to thermal (i.e.\ that of a blackbody).  Heat radiating from a surface of a given area with a given flux has the highest entropy when the spectrum is thermal, so a non-thermal spectrum for that energy would imply that free energy remains to be tapped from the energy at constant surface area.  A K2 optimizing its operating temperature to balance engineering difficulty with thermodynamic efficiency would thus aim for its waste heat to have an SED as close to that of a blackbody as possible.  

Thus, while an alien civilization using starlight might emit waste heat over a range of temperatures, we should not expect it to be so cold or to have such a strongly non-thermal spectrum that as to be easily missed in the MIR.

\subsubsection{Other Assumptions: Steady State, and Conservation of Energy}

More advanced civilizations might do significant work that efficiently stores energy as kinetic or potential energy, rearranging the mass of their stellar system or galaxy in both position and velocity space, and violating our steady-state assumption.  Such a process would be inherently limited, however, as eventually they must either ``evaporate'' their stellar system or galaxy, or else heat it until it radiates thermal waste heat.  

More exotically, if the ``dark sector'' (dark energy and dark matter) contains energy that can be tapped (as a source for $\epsilon$), then it might also serve as a sink for waste heat, implying that both energy generation and disposal might be performed ``for free'' and have no detectable effects.  If such physics and engineering is ``obvious'' in the sense that advanced civilizations inevitably discover and use it, then waste heat might be only a trace component of most advanced civilizations' energy budgets.   If such engineering is so ``obvious'' and practical as to be nearly universally employed by ETI's, then a waste heat search would fail.

\section{The \ghat\ Search for Extraterrestrial Civilizations with Large Energy Supplies}

\label{WISE}

We have begun the G-HAT search for alien waste heat, which we denote \ghat.  We will give detailed descriptions of our search methodology in later papers, but here we will give a general outline.  We also discuss some of the most promising ways one could remove false positives by identifying dust as the source of infrared radiation identified in a broadband MIR survey.  We also show give a ``zeroth'' order result of our search from our preliminary analysis of the {\it WISE} all-sky survey.

\subsection{Predicted Magnitudes and Colors in {\it WISE} Bands}

The simple blackbody model of Section~\ref{Toy} is obviously an insufficient description of the actual characteristics of a galactic SED, in that both the stellar population and the dust responsible for the MIR emission cannot really be characterized by a single pair of temperatures.  The assumption of a single temperature for the stellar photosphere is more appropriate for searches of incomplete Dyson spheres, but in this case the value derived for $\alpha$ is likely to be incorrect, both because backwarming will make the MIR color temperature of the star inconsistent with its effective temperature, and because of the incorrect assumption in the $(1-\alpha)$ term that we can effectively model the absorption of starlight as a translucent screen (see \S~\ref{spectra}).

For galaxies, we can do better by employing galaxy templates\footnote{We obtained these templates from the SWIRE template library at http://www.iasf-milano.inaf.it/$\sim$polletta/templates/swire\_templates.html, compiled and hosted by Mari Polletta} generated by Mari Polletta using the GRASIL code of \citet{Silva98}, to provide a more realistic functional form for $f\host$. Inspection of the elliptical and spiral galaxy templates reveals an appropriate value for $T_*$ in the toy model would be $T_*=4000$K, consistent with most of the starlight in a typical galaxy, especially in the red optical, coming from K giants.  

To transform the predictions of the AGENT formalism and the Silva SEDs into predictions of colors in the \WISE\ bands, we followed the prescription in the \WISE\ Explanatory Supplement\footnote{http://wise2.ipac.caltech.edu/docs/release/allsky/expsup/index.html} and references therein \citep{WISEexp,WISE,WISEphot}.  

Using the filter response curves and zero points described in \citet{WISEexp}, we constructed a procedure that calculates the received flux in the four bandpasses of \WISE\ for a blackbody emitter at temperature \twaste\ with fiducial luminosity 1 \lsun\ at a distance of 10 pc. Rather than compute the response to a blackbody anew every time we need them, we fit the resulting fluxes in each of the four bandpasses as a function of temperature with the function $\sum_{i=0}^{3}= a_i \log{\twaste}^{i-1}$.  Use of this simple expression in $T$ significantly speeds computation time.  Our procedure then scales these predicted fluxes by the luminosity and distance of a given hypothetical source.  

To validate our calculations, we used this procedure to calculate \WISE\ magnitudes of the star Vega (since \WISE\ uses Vega magnitudes).  This procedure was complicated by the fact that Vega is not a blackbody, but in the Rayleigh-Jeans limit in the MIR this is a reasonable approximation.  We consulted \citet[][Fig. 2]{Vega} to determine the appropriate value of $T_*$ to use, and found that at 5$\mu$, a continuum temperature of 13,000 K produced the best match to the spectra in that work.  Our procedure predicts $(W2-W1)$, $(W3-W1)$, and $(W4-W1)$ colors of $(-0.0011, -0.0002, 0.020)$ for blackbodies at this temperature.  These values suggest that our approximation of Vega as a blackbody induces errors of $<1\%$ in the first two colors, and of order 2\% in $(W4-W1)$; this serves as a reality check on our interpretation of the {\it WISE} response curves and zero points.  

We apply this computation to the K3 hunt by using this procedure to calculate the sensitivity of \WISE\ to objects with the spectra described in Eq.~\ref{alphaeq} with $\alpha=\gamma$ and $\nu=0$.  Specifically, our procedure calculates \WISE\ fluxes for the old elliptical {\it Ell13} Silva template and blackbodies of temperature \twaste\, and scales them by $(1-\alpha)$ and $\gamma$, respectively.  It then sums these two fluxes in each of the four \WISE\ bands, and scales them appropriately by $(L\host/(4\pi d^2))$ to produce the predicted \WISE\ magnitudes for a a given set of $\alpha, \gamma, \twaste, (L\host/(4\pi d^2))$.  

\subsection{Interpreting \WISE\ SEDs with AGENT Parameters}

\label{interpret}

We now give a few illustrative examples in the context of our AGENT parameterization from Eq.~\ref{alphaeq}  and sources in the \WISE\ Source Catalog.
 
To interpret the observed \WISE\ magnitudes of sources in terms of our formalism, we inverted our modeling process assuming $\alpha=\gamma$ (thus the model has three free parameters: $\gamma, \twaste$, and $(L\host/(4\pi d^2))$).  In cases where \WISE\ provides only an upper flux limit in two bands, the model can be underconstrained.  We found that our fitter (based on the Levenburg-Marquardt algorithm in MPFIT \citep{Markwardt09,LMfit}) converged in the vast majority of cases, and that the typical failure modes were in cases where the observed magnitudes were strongly inconsistent with any positive parameter values in our model.  

\begin{figure*}
\epsscale{0.84}
\plotone{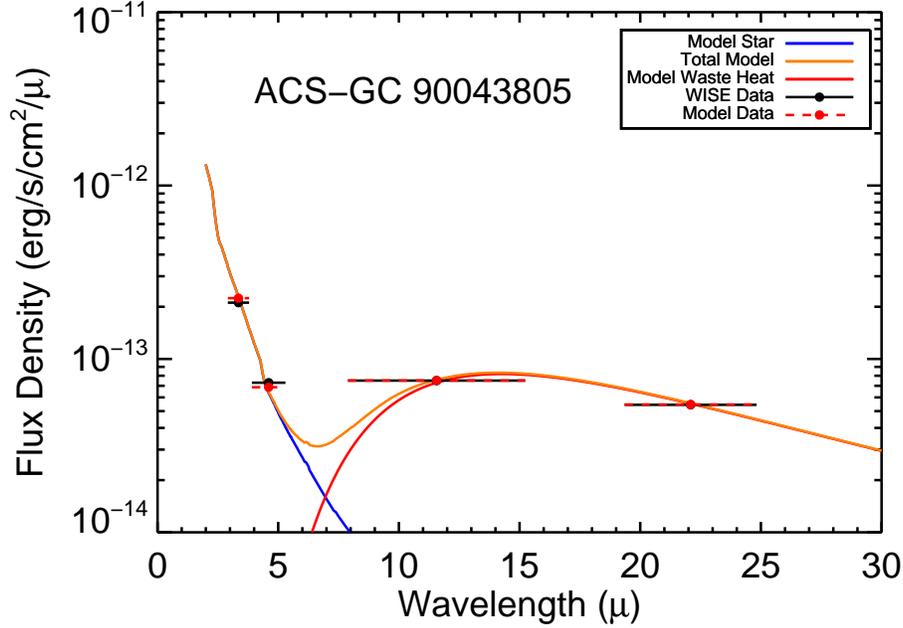}
\caption{Model and observed MIR SED of spiral galaxy ACS-GC 90043805 \citep{Griffith12}.  Black points represent observed \WISE\ magnitudes and the corresponding horizontal line corresponds to the approximate span of the waveband.  The orange line is the best-fit model SED of our toy model from Eq.~\ref{alphaeq} with AGENT parameters $\alpha=\gamma=$ 0.20, and \twaste=200 K.  The red points with dashed lines correspond to the {\it model} magnitudes in the \WISE\ bands of the model SED (which are coincident with the black points and lines in this case because the fit is essentially perfect).  This typical example of a spiral galaxy illustrates why a search for waste heat from K3's with broad band photometry will necessarily be limited to civilizations with values of $\gamma$ significantly higher than this because of confusion with ordinary galaxies. \label{Spiral}}
\end{figure*}

In Figure~\ref{Spiral} we show the SED and model fit spiral galaxy to illustrate how our toy model characterizes the stars and dust there.  If we were to interpret this as an old elliptical galaxy full of civilizations, we would say that $\gamma = 20\%$, meaning that 20\% of the starlight has been reprocessed to a low waste heat temperature, $\twaste=200$ K.  

Naturally, the true interpretation of this galaxy's SED involves both dust and stellar photospheres with a range of temperatures.  A more secure value of $\gamma$ (interpreted more naturalistically as the amount of starlight reprocessed by dust) would require comparing the luminosity of the dust emission (primarily in the FIR) to that from unobscured starlight.  A more rigorous approach to the problem would be to use the dust models of \citet{Draine07} to fit available photometry and in terms of the physically motivated parameters in that model: $q_{\rm PAH}$, $\gamma_{\rm Draine}$, $M_{\rm dust}$ and $U_{\rm min}$, representing the fraction of dust in polycyclic aromatic hydrocarbons (PAHs), the fraction of dust illuminated by starlight with intensity $U>U_{\rm min}$, and the total dust mass in the galaxy.  Interpreted with these parameters, even broadband photometry of a K3, given enough bands, might result in unusually large or even unphysical values of $q_{\rm PAH}$ (since the excess 12$\mu$ emission from the waste heat would be interpreted in this model as excess emission from PAH's).

The example in Figure~\ref{Spiral} is at the high end of the range of $\gamma$ values we derive for typical galaxies, and very few sources are consistent with $\gamma>0.4$.  This means that efficient, broadband searches for waste heat from K3's will necessarily be restricted to $\gamma$ values significantly higher than this, where few astrophysical sources are to be found.

\begin{figure*}
\epsscale{0.84}
\plotone{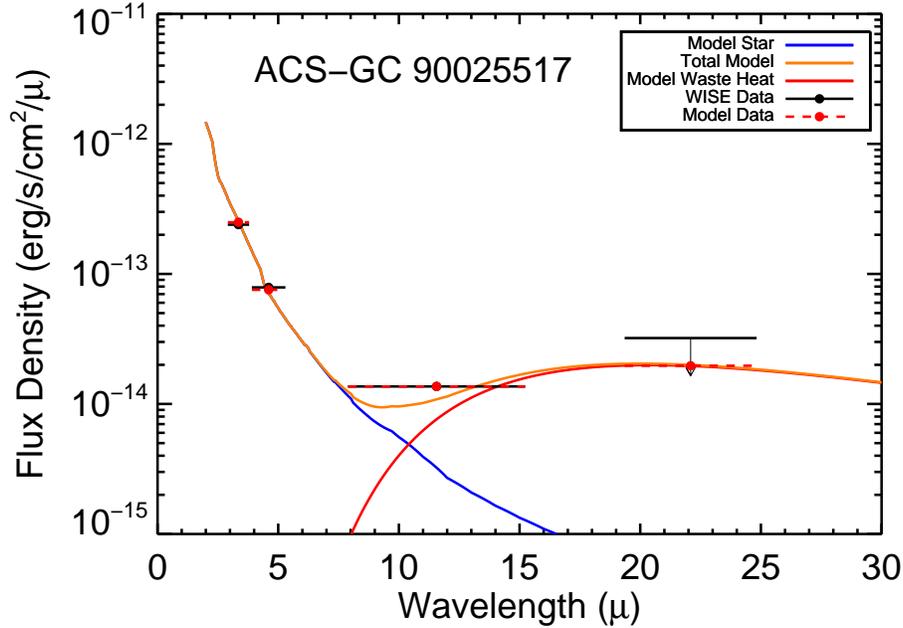}
\caption{Model SED and observed MIR SED of lenticular galaxy ACS-GC 90025517 \citep{Griffith12}.  Black points represent observed \WISE\ magnitudes and the corresponding horizontal line corresponds to the approximate span of the waveband, including a non-detection (upper limit) in $W4$.  The orange line is the best-fit model SED of our toy model from Eq.~\ref{alphaeq} with AGENT parameters $\alpha=\gamma=$ 0.07, \twaste=142 K.  The red points with dashed lines correspond to the {\it model} magnitudes in the \WISE\ bands of the model SED.  This example illustrates how the model fitter properly handles upper limits and how galaxies with less dust than typical spirals yield small $\gamma$ values. \label{S0}}
\end{figure*}

Figure~\ref{S0} illustrates a lenticular galaxy, which, having less gas than a typical spiral, yields a correspondingly lower value of $\gamma = 0.07$.  This case also illustrates how the fitter can handle upper limits properly.  Still, it is clear that, despite (or because of) the non-detection in $W4$ for this galaxy, high $\gamma$ values are inconsistent with the data, and we can put an upper limit on the size of the energy supply of any K3 civilization in it of 7\% of its starlight.


The assumption that $\alpha=\gamma$ means that our interpretation of $\gamma$ may be incorrect if, in fact, the civilization primarily uses non-stellar sources for its energy supply (see Section~\ref{Esources}).  If, in fact, $\alpha\sim0$, then the correct value of $\gamma$ will be related our inferred value, $\gamma^\prime$ by $\gamma = \gamma^\prime/(1-\gamma^\prime)$.  Thus, if we measure $\gamma^\prime=0.9$ assuming $\alpha=\gamma$, then the corresponding value in the $\alpha=0$ case is $\gamma = 0.9/0.1 = 9$, meaning that the civilization has an energy supply that is nine times larger than the stellar luminosity available to it.  

Put another way, our fitting procedure would measure a K3 with $\epsilon = \gamma >1$ (more than a galactic luminosity of non-stellar energy supply) to have $\gamma^\prime > 0.5$ (because we improperly assumed $\alpha=\gamma$).  This is above the value we derive for most galaxies, so \WISE\ is capable of effectively ruling out the existence of any galaxies with $\epsilon >1$.

\subsection{Contamination From Natural Sources} 

The process of identifying alien civilizations will necessarily be one of exclusion of natural sources, at least until an unambiguously alien signal is identified (\S~2.3 of Paper I).   The characteristics of waste heat of high luminosity are that it will be bright and have a low color temperature.  This is also a characteristic of astrophysical dust, which will be a major source of false positives in any waste heat SETI work.  This is because dust performs essentially the same task as alien civilizations would:  intercepting starlight and reradiating it in the MIR (although in the case of dust, the radiation is often a non-thermal process producing characteristic emission from polycyclic aromatic hydrocarbons).

A broadband search cannot distinguish the spectral features of dust from that of other sources of MIR radiation.  This means that without additional information (such as spectra or contextual information like an object's location in a star forming region) civilizations with low levels of waste heat will not be distinguishable from ordinary astrophysical sources.  The most populous class of natural sources with the highest MIR to near-infrared flux ratios will set a lower limit to the energy supply of extraterrestrial civilizations we are sensitive to, and we can identify galaxies and stars with MIR fluxes much higher than this limit for follow-up.

\begin{figure*}
\iftoggle{emulateapj}{
  \hspace{-0.8in}
  \includegraphics[width=3in,angle=270]{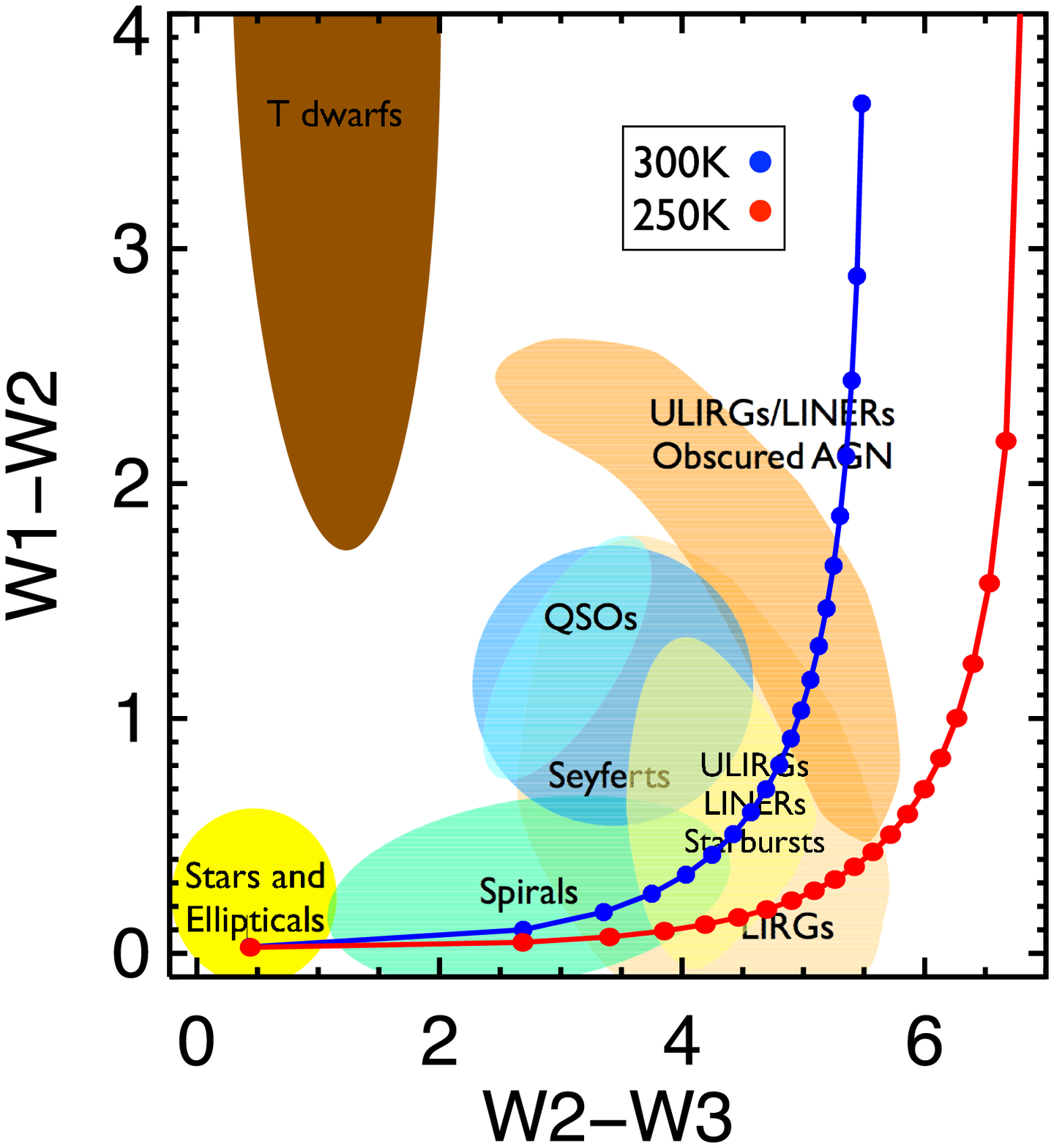}
  \includegraphics[width=3in,angle=270]{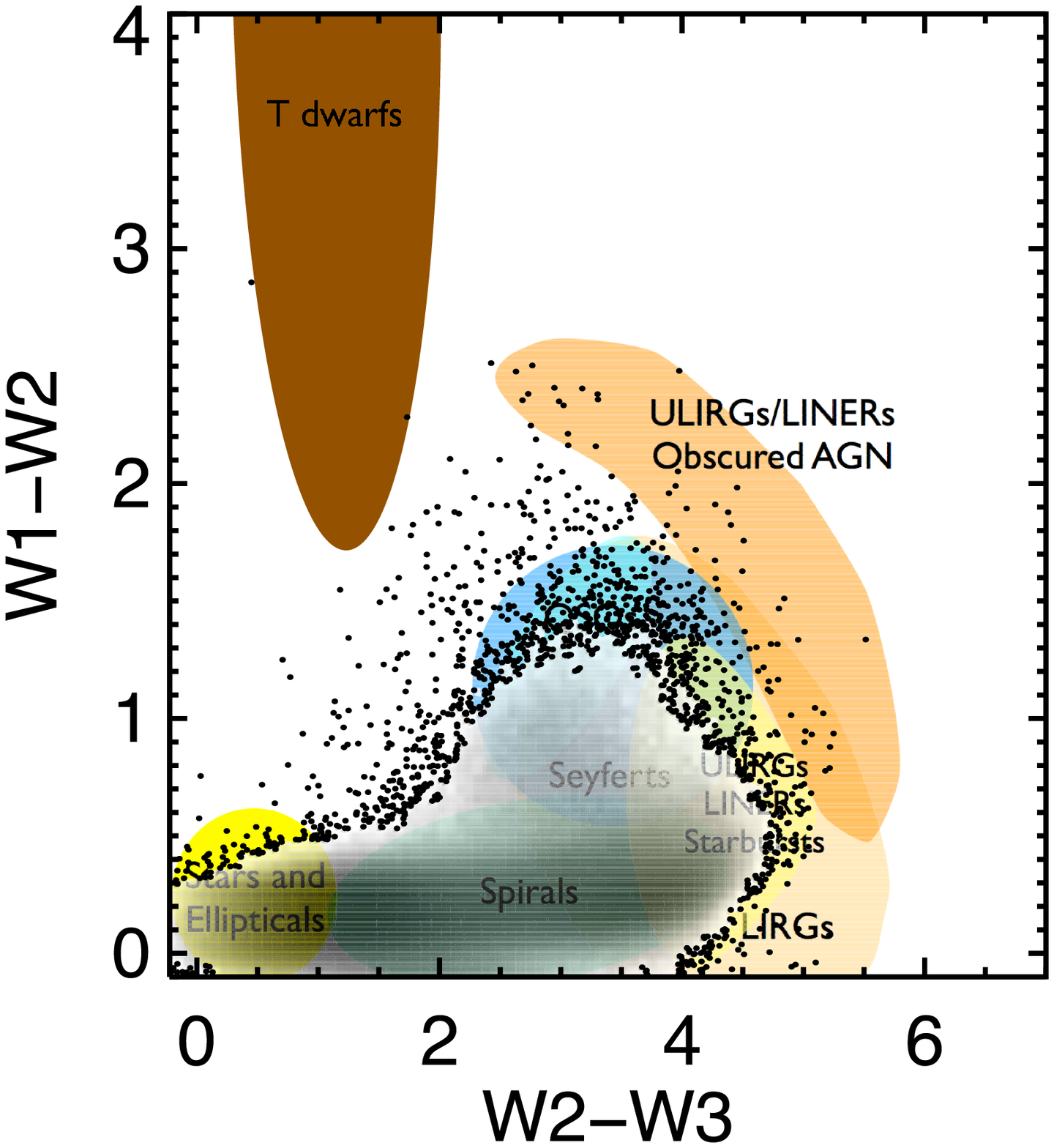}
  \vspace{0.4in}
}{
  \plotfiddle{cc1}{0pt}{270}{250}{330}{-110}{0}  
  \vspace{-290pt}
  \plotfiddle{cc2}{0pt}{270}{250}{330}{200}{550}
}
\caption{Color-color space of some natural and artificial sources in the \WISE\ bands.  We recreated the underlying figures from \citet{Lake12} and \citet{WISEphot} {\it Left:} The colors of K3 civilizations with AGENT parameters $\alpha=\gamma$, $\epsilon=\nu=0$ and two values of \twaste.  Values for $\gamma$ increase from zero (lower left) to 1.0 in increments of 0.05.  Civilizations with high values of $\gamma$ are distinct from regions populated by most known classes of sources.  {\it Right:} Distribution of sources listed in both the \WISE\ Source Catalog and the 2MASS Extended Source Catalog with good point source photometry in \WISE\ (so the vast majority of these are galaxies with good colors).  In regions of high source density, we have used a logarithmic greyscale.  The region of color space for K3's with $\gamma>0.4$ is largely free of contamination, illustrating how \WISE\ provides good discriminatory power between artificial sources and most natural extragalactic sources, with only ULIRGs and similarly dusty galaxies providing confusion.\label{cc}  }
\end{figure*}

Figure~\ref{cc} shows the distribution of extragalactic sources in the \WISE\ Source Catalog.  We generated this sample by cross-matching the \WISE\ Source Catalog with the 2MASS Extended Source Catalog to remove stars.  To ensure that the photometry is robust for this figure, we excluded sources resolved by \WISE\ (using the filter \verb#EXT_FLG == 0#) and used the profile-fitted photometry (our full search will include resolved sources, whose nature will be easier to diagnose than point sources).  While some of these sources may be Galactic, the vast majority of \WISE\ sources are extragalactic \citep{WISEexp}.  We recreated the underlying figure from the works of \citet{Lake12} and \citet{WISEphot}, who have helpfully identified regions in this color space where many classes of natural extragalactic objects reside.  

Figure~\ref{cc} shows that artificial sources with $\gamma>0.4$ occupy a region of color space beyond most classes of natural, extended sources, and indeed that nearly all of the sources detected by \WISE\ are consistent with $\gamma<0.3$.  This illustrates that the dominant source of confusion in a K3 search with \WISE\ will be LIRGs, ULIRGs, and hyper-luminous infrared galaxies, and that the MIR colors of K3 host galaxies with even modest values of $\gamma$ will be easily distinguished from those of normal galaxies.  

\subsection{Distinguishing Waste Heat in K3's (Galaxy-Spanning Civilizations) from Other Sources}

\label{K3vet}

K3's would have an additional characteristic making them easily distinguished from most forms of contamination:  they would be extended sources, even to a mission such as {\it WISE}, out to great distances.   This eliminates most forms of false positives above the Galactic plane.  Even galaxies that appear unresolved to \WISE\ will be identifiable as such if their values of $\alpha$ are sufficiently low (and their optical/NIR luminosities sufficiently high) that they appear in shorter-wavelength surveys of sufficient resolution and sensitivity \citep[for instance, in the 2MASS Extended Source Catalog,][]{2MASSXSC}.

The dominant contaminants for a search of K3's will be diffuse emission from warm dust and gas in the interstellar medium (perhaps warmed by embedded objects, such as protostars), planetary nebulae and their precursors, dusty resolved galaxies (starbursts and ULIRGs), and blue compact dwarf galaxies \citep[e.g.][]{Griffith11}.  Many of these MIR-bright sources have been cataloged already, or are only expected in regions of known star formation or dust.  Planetary nebulae are particularly easily distinguished by their characteristic optical emission and morphology.    The dominant source of contamination from extragalactic sources will thus be from ULIRGs, and perhaps galaxies with active nuclei for systems not well resolved by \WISE.

\begin{figure*}
\iftoggle{emulateapj}{
  \hspace{-0.6in}
  \includegraphics[width=5in,angle=270]{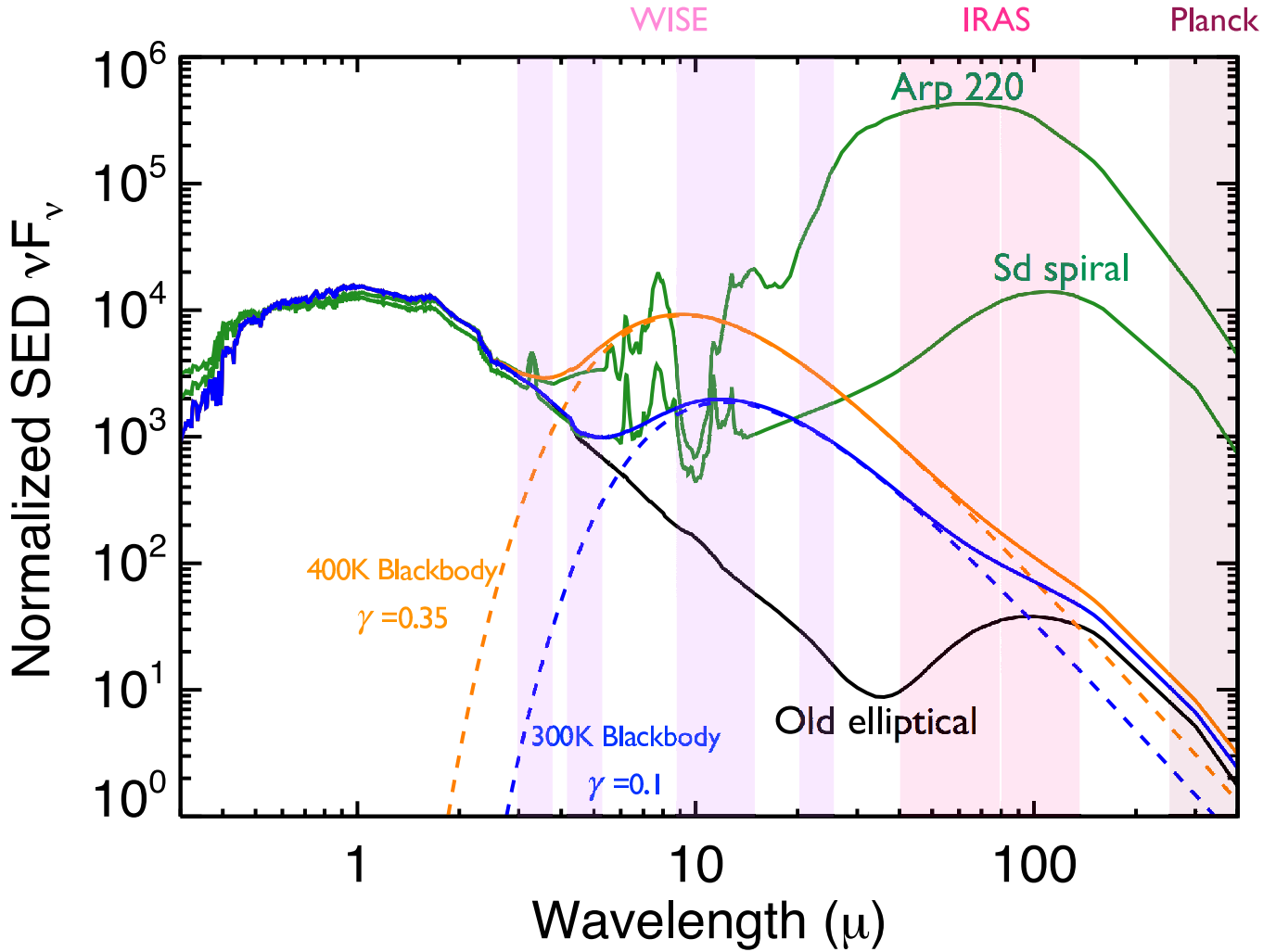}
  \vspace{0.6in}
}{
  \plotfiddle{SilvaFig}{0pt}{270}{340}{440}{-30}{-50}  
}
\caption{SEDs of typical galaxies containing dust and Kardashev civilizations, with the four {\it WISE} bands, the {\it IRAS} 60 and 100 $\mu$ bands, and the {\it Planck} 857 GHz (350 $\mu$) bandpasses highlighted.  The Galaxy template SEDs are from \citet{Silva98} and are normalized at 0.55 $\mu$.  Black: the ``old elliptical'' template {\it Ell13}.  Green: the {\it Sd} template, representing a typical late spiral, and {\it Arp220}, representing a ULIRG/starburst galaxy.  Dashed lines represent blackbodies of the listed temperatures, scaled with respect to the old elliptical template according to the AGENT parameterization with $\gamma=\alpha$ and the value of $\gamma$ indicated on the plot.  Solid orange and blue lines represent the model spectra of an galaxy composed solely of stars characteristic of an old elliptical and a K3 civilization with the AGENT parameters listed in the labels of the dashed curves of corresponding color.  The values of \twaste\ and $\gamma$ were chosen so that the resulting SEDs would roughly match the MIR flux 
observed by {\it WISE} for the spiral and starburst templates, to illustrate the degeneracy between PAH emission from dust and waste heat from ETIs.  This figure also shows how MIR spectra and FIR or sub-mm photometry might break this degeneracy.  Since even extremely dusty galaxies as Arp 220 yield derived $\gamma$ values $\sim 0.35$, K3 civilizations with even higher values of $\gamma$ would stand out as having $W2$-$W3$ colors even more extreme than these superlative objects.   
\label{Silva}}
\end{figure*}
Figure~\ref{Silva} illustrates the effects of dust on a galaxy SED, with some relevant all-sky survey bandpasses highlighted.  We have chosen three template spectra from \citet{Silva98} to span the extremes of dust content in galaxies: the ``Ell13'' ``old elliptical'' template, having almost no dust, the ``Sd'' template having dust typical of a late spiral, and a template for Arp 220, the prototypical ULIRG / starbust galaxy.  While even a small ($\sim 1\%$) midinfrared excess from a K3 civilization would be detectable by {\it WISE}, it would not necessarily be distinguishable from a small amount of dust.  Indeed, a typical Sd spiral has MIR luminosities comparable to a $\gamma=\alpha\sim 0.1$ K3 civilization in an otherwise dust-free galaxy, and values of $\gamma$ as high as 0.4 might be consistent with ULIRGs.  However higher values of $\gamma$ would be consistent only with extraordinary levels of dust and star formation (which are astrophysically interesting in their own right).  A search for extended sources with very red $W2$-$W3$ colors would be a search for high-$\gamma$ objects in a region of parameter space relatively clear of naturally red objects.

There are several ways to distinguish dust from artificially generated waste heat in a galaxy, and Figure~\ref{Silva} illustrates two of them.  

\begin{itemize}
\item Empirically, PAH emission is well correlated with [\ion{C}{2}] emission \citep{Helou01}, so an anomalously low [\ion{C}{2}]/MIR flux ratio might indicate a non-PAH origin of a galaxy's MIR luminosity.
\item MIR spectroscopy will detect the characteristic PAH emission and silicate absorption features from dust, while one would not expect those features to be present if the origin of the bulk of the MIR radiation were thermal waste heat.   Depending on the brightness of the source, CANARICAM on the Gran Telescopio Canarias, {\it SOFIA}, or the {\it James Webb Space Telescope} might be best suited to perform this followup.  
\item If PAH emission is ruled out, a rarer potential confounding source of MIR radiation is continuum emission from warm dust from high star formation rates in low-metallicty galaxies, for instance, from blue compact dwarfs \citep{Sajina07,Galliano03,Griffith11}.  In these cases, other tracers of star formation must be sought out, for instance, the MIR emission from ionic species such as [\ion{S}{2}], [\ion{Ne}{2}], or [\ion{O}{4}], which may be diagnostic of both the hot gas that accompanies the star formation and the metallicity of the galaxy.
\item Dust radiates primarily in the FIR, while artificial waste heat should radiate primarily in the MIR owing to its higher effective temperature (Section~\ref{twaste}).  The FIR luminosities of the two scenarios should thus differ by around two orders of magnitude for a given MIR luminosity, in principle making $L_{\rm FIR}/L_{\rm MIR}$ or $L_{\rm sub-mm}/L_{\rm MIR}$ a useful discriminant.  Instruments such as SCUBA-2 on the James Clerk Maxwell Telescope might be used for such a purpose.
\item Since the presence of dust implies molecules, the total mass of CO should be commensurate with dust emission, and thus diagnostic of the PAH origin of MIR emission.  Empirically, \citet{Pope13} find a slightly sub-linear relation between the 6.2$\mu$ MIR luminosity $L_{\rm PAH}$ and the line luminosity $L^\prime_{{\rm CO}(1-0)}$ for ULIRGS in both the local universe and at redshift.   From a theoretical angle, the models of \citet{Draine07} would provide a prediction of the range of expected total CO mass for a given MIR SED.  In any case, an extremely high ratios of the form $L_{\rm MIR}/L^\prime_{\rm CO}$ are inconsistent with a dusty galaxy and therefore diagnostic of artificial waste heat. 
\end{itemize}

Thus, K3's may be ultimately easier to diagnose and distinguish from natural sources than K2's, because the number of classes of potentially confounding sources is smaller than for point sources, because these sources in these classes are rarer, and because there are clear diagnostics that distinguish them.

Indeed, explorations of the dust properties of MIR-bright galaxies for the purposes of studying the relationship between dust, metallicity, and star formation, might inadvertently identify good K3 candidates as those appearing to show anomalously high PAH mass fractions \citep[e.g.\ ][]{Galametz09}. 

\subsection{Debris Disks}

An even more pernicious contaminant than dust will be stellar debris disks.  The SED's of debris disks are generally characterized as ordinary stars with some fraction of their light reprocessed by dust and larger objects, which reradiate as a roughly thermal spectrum with characteristic temperatures of tens to hundreds of Kelvins \citep{Wyatt2008}.  Indeed, the parameters often used to describe debris disk SED's, $f$ and $T$, map very will to the AGENT parameters, with $f = L_{\rm IR} / L_*$ being equivalent to our $\gamma$, and $T=\twaste$.  Debris disks are, in fact, defined by $f (= \gamma) < 10^{-2} $, making them a primary confounder for sensitive searches to Dyson spheres.  

Indeed, a broadband (or even spectral) search for waste heat from K2's is functionally equivalent to a search for debris disks.  Distinguishing Dyson spheres from debris disks might involve high-resolution imaging to show that the waste heat is not confined to a disk, or studies of the star itself to show that it is old (since debris disks are generally associated with younger ($< 1$ Gyr) stars).

Put another way, studies of and searches for debris disks are effectively searches for Dyson spheres, as well.  Of course, as scientists, we must always reach first for the naturalistic explanation for observations (see Section 2.3 of Paper I), and there is, at present, no reason to believe that stars known to have debris disks are, in fact, K2's.  Nonetheless, it would be reasonable for communication SETI efforts to make a higher priority of stars known (or rather, believed) to host debris disks, since if K2's with large energy supplies do exist in the Milky Way they would naturally appear on lists of such stars.

\subsection{Distinguishing Waste Heat in K2's (Dyson Spheres) from Other Sources}

K2's will appear as point sources, and so potential sources of confusion come from dusty stars, heavily extinguished stars, and potentially from unresolved dusty galaxies, dusty galactic nuclei, merging galaxies, or redshifted cosmological sources.  

The primary exclusion of embedded sources and other heavily extinguished sources would come from our knowledge of the distribution of dust and star formation in the Galaxy, which allows us to mask out regions where such sources are expected.  Young stars with disks will show MIR excess even outside of dusty lines of sight and will put a lower limit on the values of $\gamma$ one can search for without additional follow-up observations to determine if the star is sufficiently young to host an MIR-bright disk.

After disks, asymptotic giant stars and cosmological sources would be the other dominant contaminant in regions of the sky free from dust for high $\gamma$ candidates.  For instance, ``Hyper-luminous infrared galaxies'' \citep{HyperLIRGs}, dusty obscured galaxies \citep[``DOGs''][]{DOGs}, and similar galaxies will show very red $W2$-$W3$ colors, and may even be $W2$ dropouts.   While a number of follow-up observations could be done to identify these sources, an efficient and definitive exclusion criterion for them would be the measurement of parallax to determine their luminosity.  In particular, candidate objects with sufficient optical brightness will have their parallaxes measured with the upcoming {\it GAIA} mission \citep{GAIA}.  This would allow for candidate point sources to be segregated into three classes based on their luminosity.

\begin{itemize}
\item {\it Sources with no detectable parallax}.  These sources would be either Galactic but beyond the parallax precision of {\it GAIA}, or galaxies.  Follow-up observations at high angular resolution will distinguish these cases.  Extragalactic sources can then follow the vetting procedure described in Section~\ref{K3vet}, while Galactic sources would then have lower limits set on their luminosity that would make them consistent with high luminosity (presumably giant) stars.
\item {\it High luminosity Galactic sources}. These would most likely be AGB stars, although in principle a K2 civilization might temporarily increase the luminosity of its host star and civilization through high values of $\epsilon$.  Given that a naturalistic explanation exists for these, follow up of these objects might be of low priority, but could proceed with MIR spectroscopy following the procedure of \citet{carrigan09b}.  
\item {\it Low ($\sim$ solar) luminosity Galactic sources}. A high MIR--NIR flux ratio for an object with apparently near solar luminosity would indicate significant circumstellar dust around a dwarf star.  This could be diagnosed by identifying signs of stellar youth (for instance with X rays, or the star's presence in a star forming region).  Objects that survive this vetting will be excellent candidate K2's.
\end{itemize}

The precise parallax survey of {\it GAIA} will thus be a powerful tool for identifying K2 candidates from the many bright point sources in an MIR survey.

\subsection{A Zeroth Order Result for K3's, and Prospects for Tighter Limits}

Figure~\ref{cc} gives the zeroth-order result of the \WISE\ search for K3's:  there are very few, if any, galaxies detected by \WISE\ hosting K3's with $\gamma >0.4$.  We will scrutinize the few catalog entries consistent with such civilizations to determine if they are real and have reliable photometry.  Those that are will be good candidates for follow-up to determine if they are some class of known dusty galaxy, a new class of astrophysical object, or good candidates for being a K3.  

We note that for the galaxies less luminous than the Milky Way, this upper limit corresponds to $\gamma \lesssim 4 \times 10^{10}$ \lsun, which is less than Kardashev's original energy supply threshold for a \kthree\ civilization.  Except in the cases of galaxies significantly more luminous than the Milky Way (where the starlight might overwhelm the K3 civilization), our zeroth-order result is that Kardashev \kthree\ civilizations are very rare, and most galaxies do not host them.\footnote{Or they universally lack waste heat detectable by \WISE.}  A thorough analysis of the few sources with colors consistent with such civilizations will allow us to rule them out entirely out to a significant distance, or identify the best candidates.

In addition to finding that ``classical'' \kthree\ civilizations are very rare or non-existent, this result also implies that that physics may not allow for an easily tapped source of ``free'' energy (e.g.\ zero-point energy) at high effective temperatures (see Paper I).  If such physics existed and was ``obvious'', then alien civilizations would have unlimited free energy, and would not be bound by any resource limitation (since, as we argued in Paper I, all such limitations can be overcome through application of energy).  Such a civilization could easily span its galaxy and achieve $\gamma \gg 1$ (which, since our methodology assumes $\alpha=\gamma$, we would measure as $\gamma \gg 0.5$, see Section~\ref{interpret}).  However, our zero-th order result shows that this has never happened in our search volume.  This would seem to make the likelihood that such physics exists smaller than it was before our survey.

Objects with apparent values of $\gamma >0.25$ are more numerous, but depending on their natures, they may be easily found to be inconsistent with a K3.  This may be because they are morphologically inconsistent (the MIR emission being obviously confined to dusty lanes or the nucleus of the galaxy; see Section 5.3 of Paper I) or because the object is obviously not a galaxy (e.g., a planetary nebula, reflection nebula, or other Galactic source).  In the first (resolved K3) phase of \ghat\ we will use these methods to establish an upper limit on $\gamma$ for the galaxies in our sample.

This pilot program could be extended to tighter upper limits on $\gamma$ in several ways.  

A detailed morphological study of galaxies extended in \WISE\ would allow one to calculate an uppler limit for $\gamma$ from sources other than interstellar dust, especially in ellipticals.  This is because interstellar dust is generally not smoothly distributed, but the stars are (and, especially in ellipticals, they are also generally well-mixed).  Since the spatial distribution of the component civilizations of a K3 should be the same as that of the stars, any line of sight should be adequate for measuring $\gamma$ due to stars alone.  Thus, the line of sight with the minimum measured $\gamma$ value would set an upper limit for the entire galaxy that may be significantly tighter than that from the integrated flux.  Use of {\it Spitzer} imaging would both tighten limits on nearby galaxies and extend the method to galaxies unresolved by \WISE.  An elliptical galaxy with an unusually high and uniform MIR surface brightness would be an especially enticing target for followup.

MIR spectra or FIR fluxes would allow one to use the properties of dust we outline in Section~\ref{K3vet} to predict the integrated MIR fluxes, and search for MIR excesses above expectations due to dust.  This method would push limits for $\gamma$ below what can be done with MIR photometry alone, and may allow values of $\gamma<0.01$ to be probed in dust-free ellipticals.

\section{Conclusions}
\label{conclusions}

Before the launch of \WISE\, the upper limits on energy supplies of galaxy-spanning civilizations in other galaxies were very weak (and, to our knowledge, had never been computed).   A zeroth order analysis of \WISE\ data allows us to establish here, for the first time, that alien MIR waste heat luminosities of the order of order  $10^{11}$ \lsun\ (Kardashev's original definition of a \kthree\ civilization) are very rare in the local universe, and implies that no level of technological advancement will allow any easy or obvious access to ``free'' energy (since apparently no alien supercivilizations are using such energy to outshine their host galaxies).

We have developed a formalism, which we refer to as the AGENT parameterization, for describing the energy budget of ETI's, and described past searches for and descriptions of ETI's in terms of it.   We have also used it to show how a large area MIR survey will enable a search for them both in our own Galaxy, and spanning other galaxies.  Such a search is feasible now with the release of the {\it WISE} survey and we have begun to perform it.   
 
The Galactic phase of our search will focus on finding point sources that are unusually red for a given line of sight (i.e.\ outside of known star forming regions).  {\it GAIA} will provide a powerful way to distinguish amongst cosmological sources (which might be red because of both redshift and dust), dusty giant stars, and abnormally cool sources with dwarf-star luminosities, characteristic of Dyson spheres.  Stars with debris disks will be a major confounder for the most sensitive searches for circumstellar ETI waste heat, and communication SETI efforts should consider enriching their target lists with stars thought to host debris disks, especially any such stars that show no signs of youth.
 
The extragalactic phase of our search will allow us to identify candidate alien civilizations with large energy supplies among $\sim 10^5$ galaxies using broadband photometry alone.   We find that, because they can be resolved, galaxy-spanning civilizations may be easier to diagnose and distinguish from natural sources than circumstellar civilizations.  

A thorough analysis of the extended sources in the \WISE\ catalogs will allow us establish an upper limit for the total waste heat of galaxy-spanning civilizations near $\gamma \sim 0.25$ (i.e.\ total MIR waste heat luminosity of one-fourth of the galaxy's stellar luminosity).   This limit will be significantly lower than the only previous search for galaxy-spanning civilizations, that of \citet{annis99b}, who found no evidence of galaxies with $\alpha \gtrsim 0.75$ (i.e.\ obscuration of three-fourth of the galaxy's stellar luminosity).  Morphological, spectral, and FIR or radio data can push this limit down further, to the point where it may begin to significantly sharpen the Fermi Paradox.  

Useful follow-up observations would include high-resolution imaging to determine source morphology, radio detection of molecules such as CO, mid-infrared spectra to detect PAH features, and far-infrared or submillimeter photometry to detect thermal dust emission.  Such observations will allow for detection of galaxy-spanning alien energy supplies below $\sim 25\%$ of stellar luminosity.  
 
We hope this first effort inspires more detailed surveys to do just that, and that positive detections of anomalous objects will inspire new astrophysics and provide candidates for communication SETI and other investigation.

\acknowledgements

This research is supported entirely by the John Templeton Foundation through its New Frontiers in Astronomy and Cosmology, administered by Don York of the University of Chicago.  We are grateful for the opportunity provided by this grant to perform this research.

The Center for Exoplanets and Habitable Worlds is supported by the Pennsylvania State University, the Eberly College of Science, and the Pennsylvania Space Grant Consortium.

This publication makes use of data products from the Wide-field Infrared Survey Explorer,  the NASA/IPAC Infrared Science Archive and the NASA/IPAC Extragalactic Database (NED), all of which are operated by the Jet Propulsion Laboratory, California Institute of Technology, under contract with or with funding from the National Aeronautics and Space Administration.  This research has also made use of NASA's Astrophysics Data System and the SIMBAD database, operated at CDS, Strasbourg, France.  We thank Erik Rosolowsky for making the binplot.pro and disp.pro routines available for our use in making Figure~\ref{cc}.

We acknowledge that many ideas not original to us are incompletely cited (often because we are not aware the earliest or most appropriate references); we apologize that the references included in this series of papers are not comprehensive.  For instance, Zubrin's loose use of the term ``\kthree\ civilization'' appears in \citet{kardashev85} but likely has earlier attestations; our use of the shorthand $K1$, $K2$, and $K3$ is at least as old as \citet{Webb2002}, but may have earlier origins.
 
We are grateful for the assistance of IPAC and members of the \WISE\ team, especially Davey Kirkpatrick, Tom Jarrett, and Michael Cushing.  We thank Arjun Dey, Eugene Chiang, and Marc Kuchner for helpful contributions and discussions.  We thank Adam Kraus for the insight that {\it GAIA} will provide a powerful discriminant for K2's.  We thank our referees for appropriately challenging our conclusions and sharpening our reasoning.

\pagebreak

\end{document}